\documentclass[12pt]{iopart}
\pdfoutput=1
\usepackage{graphicx}

\begin{document}

\title[Entanglement Spectra and Entanglement Thermodynamics of Hofstadter Bilayers]{Entanglement Spectra and Entanglement Thermodynamics of Hofstadter Bilayers}

\author{John Schliemann}

\address{Institute for Theoretical Physics, University of Regensburg,
D-93040 Regensburg, Germany}
\ead{john.schliemann@physik.uni-regensburg.de}
\begin{abstract}
We study Hofstadter bilayers, i.e. coupled hopping models on two-dimensional 
square lattices in a perpendicular magnetic field. Upon tracing out one of the
layers, we find an explicit expression for the resulting entanglement spectrum
in terms of the energy eigenvalues of the underlying monolayer system. 
For strongly coupled layers the entanglement Hamiltonian is proportional
to the  energetic Hamiltonian of the
monolayer system. The proportionality factor, however, cannot be
interpreted as the inverse thermodynamic temperature, but represents a
phenomenological temperature scale. We derive an explicit relation between
both temperature scales which is in close analogy to a standard result of
classic thermodynamics. In the limit of vanishing temperature, thermodynamic
quantities such as entropy and inner energy approach their ground-state
values, but show a fractal structure as a function of magnetic flux.
\end{abstract}

\section{Introduction}

The study of quantum entanglement has by now developed to a mature subfield of 
many body physics \cite{Amico08,Tichy11,Eisert10}. 
Among the recent developments, the
concept of the entanglement spectrum \cite{Li08} has been applied to a
plethora of different systems. 
These comprise
quantum Hall monolayers at fractional filling 
\cite{Li08,Regnault09,Zozulya09,Lauchli10,Thomale10a,Sterdyniak10,Thomale10b,Chandran11,Sterdyniak11,Liu12,Qi11,Alba11,Sterdyniak11a,Dubail11b,Rodriguez11,Sterdyniak12} 
where
various spatial geometries and ways of separating the system into subsystems
have been investigated. In the case of bilayer quantum Hall systems, a very
natural way of defining a subsystem is provided by the double layer structure.
Specifically, a numerical investigation of quantum Hall bilayers at total
filling factor of unity revealed a striking similarity between the entanglement
spectrum of the composite systems and the energy spectrum of a monolayer
\cite{Schliemann11}. A analogous observation was reported slightly earlier
in a numerical study of two-leg spin-$1/2$ ladders \cite{Poilblanc10}
 which was subsequently
understood in terms of perturbation theory in the limit of strong
rung coupling \cite{Peschel11,Lauchli11}, a result which is  
remarkably also valid for arbitrary spin length \cite{Schliemann12}.
Similar findings were obtained in the simple case of coupled chains of 
free fermions where explicit expressions for the entanglement
spectrum at arbitrary coupling can be derived \cite{Peschel11,Qi12}.
The starting point of the present work is to extend these results to Hofstadter
bilayers, i.e. coupled hopping models on two-dimensional square lattices
in a perpendicular magnetic field \cite{Hofstadter76}. 
The energy spectrum of such systems generates, as a function of the 
magnetic flux per unit cell, highly self-similar and visually appealing
structures known as Hofstadter butterflies.
As we shall see below, these features immediately translate to 
the entanglement spectrum via an explicit formula for the entanglement
levels in terms of the energy eigenvalues of the monolayer system. 
Another focus of the
present work are thermodynamic properties of the reduced density matrix
and its entanglement Hamiltonian. In particular, we derive a
thermodynamic entanglement temperature and inner energy starting from
an effective coupling parameter of the bilayer system which can be viewed 
as a phenomenological entanglement scale.

Entanglement spectra of Hofstadter monolayers were already investigated in
Ref.~\cite{Huang12} with the underlying square lattice being partitioned into
two blocks. This and related ways of defining subsystems were also used in 
other studies
of entanglement spectra of spin systems \cite{Calabrese08,Xu08,Pollmann10a,Pollmann10b,Thomale09,Franchini11,Yao10,Cirac11,Huang11,Lou11,Alba12,James12,Lundgren12,Schuch12}.
Other recent investigations in connection with entanglement spectra include
topological insulators \cite{Fidkowski10,Prodan10}, rotating Bose-Einstein 
condensates \cite{Liu11},
coupled Tomonaga-Luttinger liquids \cite{Furukawa11}, 
interacting bosons \cite{Deng11,Tanaka12,Alba12a}, 
and complex paired superfluids \cite{Dubail11a}.

This paper is organized as follows. In section \ref{nomag} we review and
extend previous results on free fermions in two-party lattice systems in the
absence of a magnetic field. In the following section \ref{entspec} we
analyze the entanglement spectrum of Hofstadter bilayers and derive
an explicit formula for the entanglement levels as a function of the
underlying (highly fractal) monolayer system. Section \ref{thermo} is
devoted to a thorough analysis of the thermodynamic properties of the
reduced density matrix and its entanglement Hamiltonian.
We close with a summary and an outlook in section \ref{concl}.

\section{Free Fermions without Magnetic Field}
\label{nomag}

Let us first briefly review and generalize results of Ref.~\cite{Peschel11}
on free fermions on coupled lattices. We consider the Hamiltonian
\begin{equation}
{\cal H}=\sum_{\vec k}\left[
\varepsilon_A(\vec k)a^+_{\vec k}a_{\vec k}
+\varepsilon_B(\vec k)b^+_{\vec k}b_{\vec k}
-t_{\perp}\left(a^+_{\vec k}b_{\vec k}+b^+_{\vec k}a_{\vec k}\right)\right]
\label{defham1}
\end{equation}
where $a^+_{\vec k}$,$a_{\vec k}$ ($b^+_{\vec k}$,$b_{\vec k}$)
generate and annihilate fermions with wave vector $\vec k$ and energy
$\varepsilon_A(\vec k)$ ($\varepsilon_B(\vec k)$) on a $d$-dimensional lattice
$A$ ($B$). The two subsystems are coupled by a hopping term proportional to
$t_{\perp}$ leading, at given wave vector, to a $2\times 2$ eigenvalue
problem whose elementary solution is
\begin{equation}
{\cal H}=\sum_{\vec k}\left[
\varepsilon_+(\vec k)\alpha^+_{\vec k}\alpha_{\vec k}
+\varepsilon_-(\vec k)\beta^+_{\vec k}\beta_{\vec k}\right]
\end{equation}
with 
\begin{equation}
\varepsilon_{\pm}(\vec k)
=\frac{1}{2}\left(\varepsilon_A(\vec k)+\varepsilon_B(\vec k)\right)
\pm\frac{1}{2}
\sqrt{4t_{\perp}^2+\left(\varepsilon_A(\vec k)-\varepsilon_B(\vec k)\right)^2}
\label{erg1}
\end{equation}
and 
\begin{eqnarray}
\alpha^+_{\vec k} & = & \eta_{\vec k}^+a^+_{\vec k}
-\frac{t_{\perp}}{|t_{\perp}|}\eta_{\vec k}^-b^+_{\vec k}\\
\beta^+_{\vec k}& = & \eta_{\vec k}^-a^+_{\vec k}
+\frac{t_{\perp}}{|t_{\perp}|}\eta_{\vec k}^+b^+_{\vec k}
\end{eqnarray}
where
\begin{equation}
\eta_{\vec k}^{\pm}=\sqrt{\frac{1}{2}\left(1\pm
\frac{\varepsilon_A(\vec k)-\varepsilon_B(\vec k)}
{\sqrt{4t_{\perp}^2+\left(\varepsilon_A(\vec k)-\varepsilon_B(\vec k)\right)^2}}
\right)}
\end{equation} 
Let us now focus on the case where the above dispersion bands do not overlap,
i.e.
\begin{equation}
\varepsilon_+(\vec k_1)\geq\varepsilon_-(\vec k_2)
\label{cond1}
\end{equation}
for all wave vectors $\vec k_1$, $\vec k_2$. This is in particular the case
if 
\begin{equation}
\left(\varepsilon_A(\vec k)+\varepsilon_B(\vec k)\right)^2
\leq\left(\varepsilon_A(\vec k)-\varepsilon_B(\vec k)\right)^2
\label{cond2}
\end{equation}
holds identically in $\vec k$ since then we have
$\varepsilon_+(\vec k_1)\geq 0\geq\varepsilon_-(\vec k_2)$ for all
$\vec k_1$, $\vec k_2$. Indeed previous work \cite{Peschel11,Qi12}
has concentrated on the
situation $\varepsilon_A(\vec k)=-\varepsilon_B(\vec k)$
where the inequality (\ref{cond2}) is obviously valid.
For a half-filled system fulfilling (\ref{cond1}), only the single-particle
states generated by $\beta^+_{\vec k}$ are occupied in the ground state,
and upon tracing out, e.g., subsystem B one obtains, following the very general
arguments given in \cite{Peschel03,Cheong04}, a reduced density
matrix of the form
\begin{equation}
\rho_{\rm red}=\frac{1}{Z}\exp(-{\cal H}_{\rm ent}) 
\label{rhored}
\end{equation}
with $Z={\rm tr}(\exp(-{\cal H}_{\rm ent}))$ and an entanglement Hamiltonian
\begin{equation}
{\cal H}_{\rm ent}=\sum_{\vec k}\xi(\vec k)a^+_{\vec k}a_{\vec k}\,.
\label{Hent}
\end{equation}
The entanglement levels $\xi(\vec k)$ are given by
\begin{eqnarray}
\xi(\vec k) & = & \ln\left(\frac{1-\left(\eta^-_{\vec k}\right)^2}
{\left(\eta^-_{\vec k}\right)^2}\right)
\label{preent1}\\
 & = &  \ln\left(
\frac{\sqrt{4t_{\perp}^2+(\varepsilon_A(\vec k)-\varepsilon_B(\vec k))^2}
+(\varepsilon_A(\vec k)-\varepsilon_B(\vec k))}
{\sqrt{4t_{\perp}^2+(\varepsilon_A(\vec k)-\varepsilon_B(\vec k))^2}
-(\varepsilon_A(\vec k)-\varepsilon_B(\vec k))}\right)\\
 & = & 2\,{\rm arsinh}\left(
\frac{\varepsilon_A(\vec k)-\varepsilon_B(\vec k)}{2|t_{\perp}|}\right)\,.
\label{ent1}
\end{eqnarray}
The exponential form (\ref{rhored}) of the reduced density matrix 
can be derived by demanding that this expression should generate
the same one-particle correlations in the monolayer subsystem as the
underlying pure bilayer ground-state density operator. Using elementary
properties of fermionic systems leads then to the explicit result
(\ref{preent1}) for the entanglement levels \cite{Peschel03}. 
We note that the entanglement spectrum  as discussed is generated by
the single-particle operator (\ref{Hent}) as appropriate for free fermions. 
This is different from the situation in interacting systems where the
eigenstates of the reduced density matrix are in general nontrivial
many-body states.

The above considerations extend the results of Peschel and Chung 
\cite{Peschel11} to arbitrary dimension $d\geq 1$ and independent dispersions
$\varepsilon_A(\vec k)\neq-\varepsilon_B(\vec k)$ in the two subsystems.
We note that the entanglement spectrum (\ref{ent1}) depends, differently from
the energy spectrum (\ref{erg1}), only on the difference
$\varepsilon_A(\vec k)-\varepsilon_B(\vec k)$, but not on both quantities
separately. Moreover, both the energy and the entanglement spectrum are
invariant under a change of sign of $t_{\perp}$.

If the two energy bands overlap. i.e. condition (\ref{cond1}) is violated,
the situation cannot be analyzed in such general terms. Here the
$\xi(\vec k)$ will in general not have support in certain parts of
the first Brillouin zone, while in other parts
two branches of entanglement levels will occur. 

\section{Hofstadter Bilayers: Entanglement Spectra}
\label{entspec}

Let us now concentrate on two coupled two-dimensional square lattices with the
Hamiltonian
\begin{equation}
{\cal H}={\cal H}_A+{\cal H}_B+{\cal H}_T
\end{equation}
where
\begin{eqnarray}
{\cal H}_A & = & -t_A\sum_{m,n}\left[
a^+_{m+1,n}a_{m,n}+a^+_{m,n+1}a_{m,n}+{\rm h.c.}\right]\,,\\
{\cal H}_B & = & -t_B\sum_{m,n}\left[
b^+_{m+1,n}b_{m,n}+b^+_{m,n+1}b_{m,n}+{\rm h.c.}\right]
\end{eqnarray}
and
\begin{equation}
{\cal H}_T=-t_{\perp}\sum_{\vec k}\left[a^+_{\vec k}b_{\vec k}+{\rm h.c.}\right]
=-t_{\perp}\sum_{m,n}\left[a^+_{m,n}b_{m,n}+{\rm h.c.}\right]\,.
\end{equation}
Here $a^{+}_{m,n}$, $b^{+}_{m,n}$ generate particles at sites $m$,$n$ in layer
$A$ and $B$, respectively., i.e. compared with equation (\ref{defham1}) we have
\begin{equation}
\varepsilon_{A/B}(\vec k)=-2t_{A/B}
\left(\cos(k_x a)+\cos(k_y a)\right)
\end{equation}
where $a$ is the lattice constant in both layers. In order to implement a
perpendicular magnetic field we work, as common, in Landau gauge with
the discretized vector potential $\vec A=(0,Bma,0)$ and concentrate on 
rational values of the magnetic flux per unit cell 
$\Phi=Ba^{2}/(h/e)=p/q$ (in units of $h/e$)
where $p$ and $q$ are integers without any
common divisor except unity. Thus, the system is periodic in the $x$-direction
with periodicity $qa$ while the periodicity in the $y$-direction is still 
the lattice constant $a$. The standard Peierls substitution \cite{Peierls33}
leads the following ansatz for the state vector,
\begin{eqnarray}
|\vec k\rangle & = & \frac{1}{\sqrt{N_xN_y}}\sum_{m,n}\Bigl[
\exp\left(ik_xma
+i\left(k_y+\frac{e}{\hbar}Bma\right)na\right)
\nonumber\\
 & & \qquad\qquad\qquad\cdot\left(\alpha_m(\vec k)a^+_{m,n}
+\beta_m(\vec k)b^+_{m,n}\right)\Bigr]
|0\rangle\,,
\end{eqnarray}
with $\alpha_{m}(\vec k)=\alpha_{m+q}(\vec k)$ and
$\beta_{m}(\vec k)=\beta_{m+q}(\vec k)$. Here $|0\rangle$ is the fermionic 
vacuum, and $N_x$, $N_y$ are the total numbers of unit cells in $x$- and
$y$-direction, respectively, assuming periodic boundary conditions 
commensurate with $q$. The stationary Schr\"odinger equation
\begin{equation}
{\cal H}|\vec k\rangle=\varepsilon(\vec k)|\vec k\rangle
\end{equation}
leads to a Harper equation being equivalent to the $2q\times 2q$ eigenvalue 
problem
\begin{equation}
H(\vec k)v(\vec k)=\varepsilon(\vec k)v(\vec k)
\label{2qeigen}
\end{equation}
for the $2q$-component spinor
\begin{equation}
v(\vec k)=\left(\alpha_0(\vec k),\dots,\alpha_{q-1}(\vec k),
\beta_0(\vec k),\dots,\beta_{q-1}(\vec k)\right)^T
\label{vecv}
\end{equation}
and
\begin{equation}
H(\vec k)=\left(
\begin{array}{cc}
t_Ah(\vec k) & -t_{\perp}{\mathbf 1}_{q\times q} \\
-t_{\perp}{\mathbf 1}_{q\times q} & t_Bh(\vec k)
\end{array}
\right)\,.
\label{hammat1}
\end{equation}
Here the matrix
\begin{equation}
h(\vec k)=
-\left(
\begin{array}{cccccc}
r_{0}(\vec k) & z(\vec k) & 0 & \cdots & 0 & z^{\ast}(\vec k) \\
z^{\ast}(\vec k) & r_{1}(\vec k) & z(\vec k) & 0 & \cdots & 0 \\
0 & z^{\ast}(\vec k) & r_{2}(\vec k) & z(\vec k) & \cdots & 0 \\
\vdots & 0 &\ddots & \ddots & \ddots & \vdots \\
\vdots & \vdots &\vdots & \ddots & \ddots & z(\vec k) \\
z(\vec k) & 0 & \cdots & 0 & z^{\ast}(\vec k) & r_{q-1}(\vec k)
\end{array}
\right)
\label{hammat2}
\end{equation}
with $r_m(\vec k)=2\cos(2\pi\Phi m+k_{y}a)$ and $z(\vec k)=\exp(ik_xa)$
corresponds to the classic Hofstadter monolayer problem \cite{Hofstadter76}.
The diagonalization of the latter quantity in general requires numerics,
and explicit analytical results are possible only for very special values
of the magnetic flux $\Phi=p/q$.

Now let the unitary matrix $u(\vec k)$ diagonalize $h(\vec k)$, i.e.
\begin{equation}
u(\vec k)h(\vec k)u^+(\vec k)=
{\rm diag}\left(\tilde\varepsilon_0(\vec k),\dots,
\tilde\varepsilon_{q-1}(\vec k)\right)\,.
\label{classhs}
\end{equation}
Since the off-diagonal elements in (\ref{hammat1}) are proportional
to the $q\times q$ unit matrix, they remain unchanged upon a simultaneous
diagonalization of both diagonal blocks. Thus, all four blocks are rendered
diagonal, and the diagonalization of (\ref{hammat1}) is reduced to 
$2\times 2$ problems of the form
\begin{equation}
\left(
\begin{array}{cc}
t_A\tilde\varepsilon_m(\vec k) & -t_{\perp} \\
-t_{\perp} & t_B\tilde\varepsilon_m(\vec k)
\end{array}
\right)
\label{hammat3}
\end{equation}
with the obvious eigenvalues 
\begin{equation}
\varepsilon_m^{\pm}(\vec k)
=\frac{1}{2}\left(t_A+t_B\right)\tilde\varepsilon_m(\vec k)
\pm\frac{1}{2}
\sqrt{4t_{\perp}^2+\left((t_A-t_B)\tilde\varepsilon_m(\vec k)\right)^2}
\label{erg2}
\end{equation}
and corresponding eigenvectors
\begin{equation}
\chi_m^+(\vec k)
=\left(
\begin{array}{c}
\eta_m^+(\vec k)\\
-\frac{t_{\perp}}{|t_{\perp}|}\eta_m^-(\vec k)
\end{array}
\right)\quad,\quad
\chi_m^-(\vec k)
=\left(
\begin{array}{c}
\eta_m^-(\vec k)\\
\frac{t_{\perp}}{|t_{\perp}|}\eta_m^+(\vec k)
\end{array}
\right)
\end{equation}
where
\begin{equation}
\eta_m^{\pm}(\vec k)=\sqrt{\frac{1}{2}\left(1\pm
\frac{(t_A-t_B)\tilde\varepsilon_m(\vec k)}
{\sqrt{4t_{\perp}^2+\left((t_A-t_B)\tilde\varepsilon_m(\vec k)\right)^2}}
\right)}\,.
\label{eta2}
\end{equation}
\begin{figure}
 \includegraphics[width=\columnwidth]{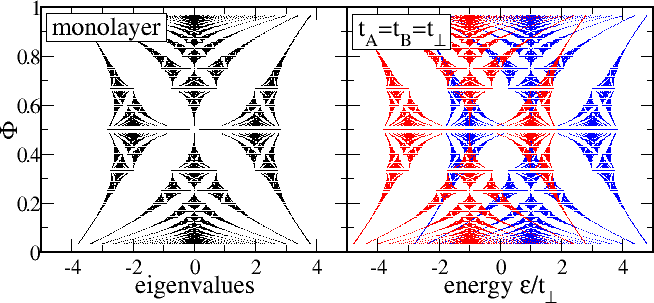}
\caption{Left panel: Classic monolayer Hofstadter spectrum (\ref{classhs})
as a function of the magnetic flux.
Right panel: Energy spectrum $\varepsilon^+_m$ (blue) and $\varepsilon^-_m$
(red) according to Eq.~(\ref{erg2}) of a bilayer system
for $t_A=t_B=t_{\perp}$. The spectrum consists of two copies 
of the Hofstadter butterfly shifted by $\pm t_{\perp}$.}
\label{fig1}
\end{figure}

Fig.~\ref{fig1} shows the numerically computed 
classic Hofstadter spectrum of a single
square lattice as a function of the magnetic flux $\Phi$ per unit cell,
along with the energy spectrum of a bilayer system
for the particularly simple case $t_A=t_B=t_{\perp}$. Here the Hamiltonian
is invariant under exchange of layers, and the eigensystem consists of states
being either symmetric or antisymmetric under this operation. Thus, as seen
in the left panel of Fig.~\ref{fig1}, the bilayer energy spectrum 
comprises two monolayer Hofstadter butterflies shifted by $\pm t_{\perp}$.

We now concentrate again on the case where both groups of dispersion
branches do not overlap,
\begin{equation}
\varepsilon_{m_1}^+(\vec k_1)\geq\varepsilon_{m_2}^-(\vec k_2)
\label{cond3}
\end{equation}
for all $\vec k_1$, $\vec k_2$ and $m_1,m_2\in\{0,\dots,q-1\}$.
Analogously to (\ref{cond2}) the stronger condition
\begin{equation}
\left(t_A+t_B\right)^2\leq\left(t_A-t_B\right)^2
\label{cond4}
\end{equation}
(i.e. $t_A$ and $t_B$ need to differ in sign) implies
$\varepsilon_{m_1}^+(\vec k_1)\geq 0\geq\varepsilon_{m_2}^-(\vec k_2)$
and therefore the inequality (\ref{cond3}).
Thus, under condition (\ref{cond3}) only the single-particle
states with energies $\varepsilon_{m}^-(\vec k)$
are occupied in the ground state of a 
half-filled system. Tracing out again layer $B$ leads to 
entanglement levels of the form 
\begin{eqnarray}
\xi_m(\vec k) & = & \ln\left(\frac{1-\left(\eta^-_m(\vec k)\right)^2}
{\left(\eta^-_m(\vec k)\right)^2}\right)\\
 & = &  \ln\left(
\frac{\sqrt{4t_{\perp}^2+((t_A-t_B)\tilde\varepsilon_m(\vec k))^2}
+(t_A-t_B)\tilde\varepsilon_m(\vec k)}
{\sqrt{4t_{\perp}^2+((t_A-t_B)\tilde\varepsilon_m(\vec k))^2}
-(t_A-t_B)\tilde\varepsilon_m(\vec k)}\right)
\label{ent2a}\\
 & = & 2\,{\rm arsinh}\left(
\frac{(t_A-t_B)\tilde\varepsilon_m(\vec k)}{2|t_{\perp}|}\right)\,.
\label{ent2}
\end{eqnarray}
These entanglement levels enter the entanglement Hamiltonian
\begin{equation}
{\cal H}_{\rm ent}=\sum_{m=0}^{q-1}\sum_{\vec k}
\xi_m(\vec k)a^+_{\vec k m}a_{\vec k m}
\label{entham}
\end{equation}
at given flux $\Phi=p/q$. Here the operators $a^+_{\vec k m}$
generate Hofstadter monolayer states with eigenvalue
$\tilde\varepsilon_m(\vec k)$, i.e. the monolayer Hamiltonian including the
magnetic field can be formulated as
\begin{equation}
{\cal H}_A=t_A\sum_{m=0}^{q-1}\sum_{\vec k}
\tilde\varepsilon_m(\vec k)a^+_{\vec k m}a_{\vec k m}\,.
\label{HA}
\end{equation}
We note that the matrix (\ref{hammat2}) has the obvious property
\begin{equation}
h(\vec k)=-h(\vec k+\vec\pi)
\end{equation}
with $\vec\pi=(\pi/a,\pi/a)$ which implies
\begin{eqnarray}
\tilde\varepsilon_m(\vec k) & = &
-\tilde\varepsilon_{q-1-m}(\vec k+\vec\pi)\,,
\label{rel1}\\
\varepsilon^{\pm}_m(\vec k) & = &
-\varepsilon^{\mp}_{q-1-m}(\vec k+\vec\pi)\,,
\label{rel2}\\
\xi_m(\vec k) & = &
-\xi_{q-1-m}(\vec k+\vec\pi)\,,
\label{rel3}
\end{eqnarray}
where we have the eigenvalues $\tilde\varepsilon_m(\vec k)$ of $h(\vec k)$
assumed to be given in ascending order. These relations enable to use, with some
caveat and specifications,
the expression (\ref{ent2}) for the entanglement levels even in situations where
the condition (\ref{cond3}) is not fulfilled:
\begin{figure}
 \includegraphics[width=\columnwidth]{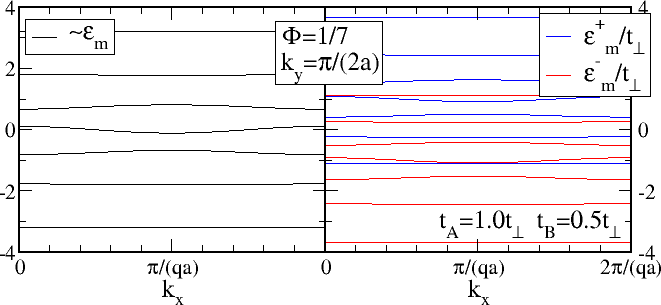}
\caption{Left panel: Eigenvalue dispersions $\tilde\varepsilon_m(\vec k)$
of the Hofstadter monolayer for $\Phi=1/7$.
Right panel: Corresponding bilayer energy dispersions $\varepsilon^+_m(\vec k)$
(blue) and $\varepsilon^-_m(\vec k)$ (red) for a choice of parameters violating
the conditions (\ref{cond3}),(\ref{cond4}).
The situation for other fluxes $\Phi=p/q$ is qualitatively similar, but with
increasing $q$ more difficult to display due to the larger number of bands.}
\label{fig2}
\end{figure}
The left panel of Fig.~\ref{fig2} shows the dispersions
$\tilde\varepsilon_m(\vec k)$ for a magnetic flux of $\Phi=1/7$ and typical
parameters. As seen in the figure, the eigenvalues form rather flat bands
which do not overlap.
The right panel displays the corresponding bilayer energy 
$\varepsilon^{\pm}_m(\vec k)$ for parameters violating the inequalities
(\ref{cond3}),(\ref{cond4}). In the ground state of a half-filled system
the lowest $q=7$ bands are completely occupied such that the highest
occupied (valence) band is $\varepsilon^+_1(\vec k)$, which is also the
highest band among the branches $\varepsilon^+_m(\vec k)$ having negative
energy. Note that $\varepsilon^+_1(\vec k)$ does also not overlap with
any other band $\varepsilon^{\pm}_m(\vec k)$; otherwise such two bands would
only partially be filled each. Let us therefore focus on this ``insulator''
scenario where
the valence band of a half filled system does not overlap with other bands
implying that in the ground state all bands are either fully occupied or 
empty. Let us now assume that some band $\varepsilon^+_m(\vec k)$
has negative energy and is therefore fully filled. Thus, according
to Eq.~(\ref{rel2}) $\varepsilon^-_{q-1-m}(\vec k)$ has positive energy
and is therefore empty.
The entanglement branch arising from $\varepsilon^+_m(\vec k)$ is
\begin{equation}
\bar\xi_m(\vec k)=\ln\left(\frac{1-\left(\eta^+_m(\vec k)\right)^2}
{\left(\eta^+_m(\vec k)\right)^2}\right)
=\xi_{q-1-m}(\vec k+\vec\pi)\,,
\end{equation}
where we have used Eqs.~(\ref{eta2}),(\ref{rel2}). Thus, up to a rigid
shift in the wave vector argument, the entanglement levels arising
from the occupied band $\varepsilon^+_m(\vec k)$ reproduce the
missing branch corresponding to $\varepsilon^-_{q-1-m}(\vec k)$. In this
sense (i.e. with the wave vector dependence being suppressed) the expression
(\ref{ent2}) for the entanglement spectrum can also be used even
if the inequality (\ref{cond3}) does not hold, but all bands are, in the
ground state, either empty or completely occupied.
On the other hand, starting directly from the original
$2q\times 2q$ eigenvalue problem (\ref{2qeigen}), the entanglement
levels can be expressed as 
\begin{equation}
\xi_m(\vec k)=\ln\left(\frac{1-\zeta_m(\vec k)}
{\zeta_m(\vec k)}\right)
\label{ent3}
\end{equation}
where
\begin{equation}
\zeta_m(\vec k)=\sum_{m'=0}^{q-1}\left|\alpha_{m'}^{(m)}(\vec k)\right|^2\,,
\end{equation}
and the $\alpha_{m'}^{(m)}(\vec k)$ are the components referring to
subsystem $A$ in the eigenvectors (\ref{vecv}) corresponding to the
lowest $q$ eigenvalues in (\ref{2qeigen}), $m\in\{0,\dots,q-1\}$.
Where applicable (see above), the results (\ref{ent2}) and (\ref{ent3})
of course coincide, as we have also checked by explicit numerical
calculations.

In summary,  the expressions (\ref{erg2}) and (\ref{ent2}) provide, 
in full analogy with Eqs.~(\ref{erg1}),(\ref{ent1}),
explicit relations between the
energy and the entanglement spectrum, respectively, of the composite
system and the energy spectrum of the monolayer.  
Moreover, the entanglement spectrum $\xi_m(\vec k)$ depends, for a given
monolayer spectrum  $\tilde\varepsilon_m(\vec k)$
only on the single quantity
\begin{equation}
\lambda:=\frac{t_A-t_B}{|t_{\perp}|}
\label{lambda}
\end{equation}
which will in the following be referred to as the (effective) coupling 
parameter.

As already discussed above, in the case $\lambda=0$
the layers are maximally entangled with each other, and
the all entanglement levels are zero.
\begin{figure}
 \includegraphics[width=\columnwidth]{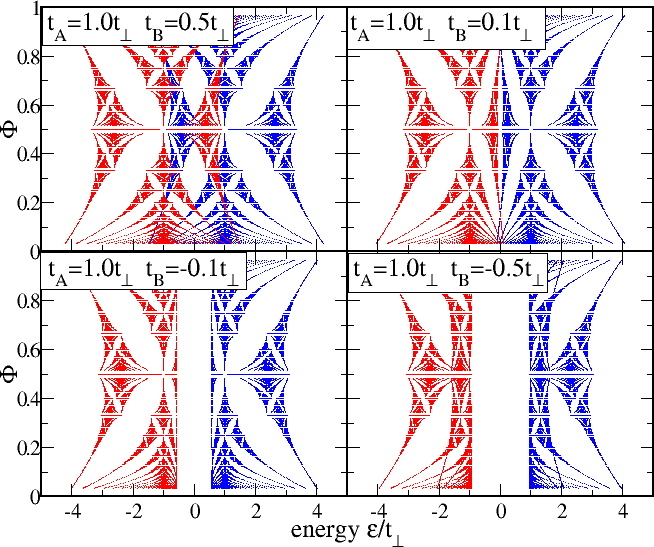}
\caption{Energy spectrum $\varepsilon^+_m$ (blue) and $\varepsilon^-_m$
(red) according to Eq.~(\ref{erg2}) of a bilayer system
for $t_A=t_{\perp}$ and various values of $t_B$.}
\label{fig3}
\end{figure}
\begin{figure}
 \includegraphics[width=\columnwidth]{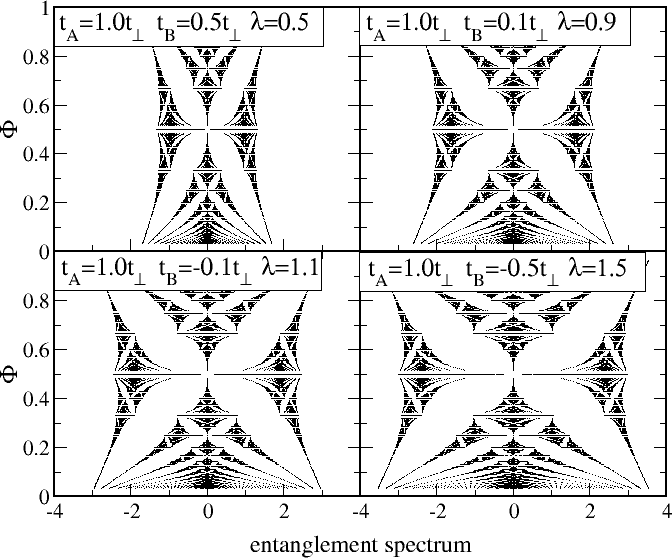}
\caption{Entanglement levels of bilayer systems
for the same parameters as in Fig.~\ref{fig3}. The spectra are Hofstadter
butterflies ``deformed'' by the inverse hyperbolic sine function 
occuring in Eq.~(\ref{ent2}). The effective coupling parameter $\lambda$
is defined in Eq.~(\ref{lambda}).}
\label{fig4}
\end{figure}
In Fig.~\ref{fig3} we have plotted the energy spectrum of the bilayer system
according to Eq.~(\ref{erg2}) for $t_A=t_{\perp}$ and various values of
$t_B$. As seen there, if $t_A$ and $t_B$ differ in sign (here: $t_B<0$),
an energy gap which is independent of the magnetic flux $\Phi$ occurs.
Fig.~\ref{fig4} displays the corresponding entanglement spectra which form 
Hofstadter butterflies ``deformed'' by the inverse hyperbolic sine function 
occuring in Eq.~(\ref{ent2}).
\begin{figure}
 \includegraphics[width=\columnwidth]{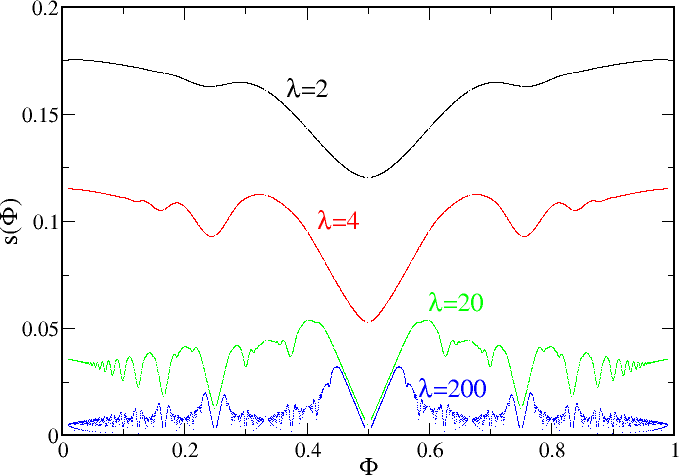}
\caption{The entropy $s(\lambda,\Phi)$ per unit cell as a function of
the magnetic flux $\Phi$ for various values of $\lambda$. For 
$\lambda\approx10$ and larger,
fractal features of $s(\lambda,\Phi)$ become visible.
The data (as well as the one in the following Figs.~\ref{fig6} and \ref{fig7})
was calculated numerically according to Eqs.~(\ref{s}),(\ref{e})
using an appropriate discretization of the magnetic Brillouin zone.}
\label{fig5}
\end{figure}
\begin{figure}
 \includegraphics[width=\columnwidth]{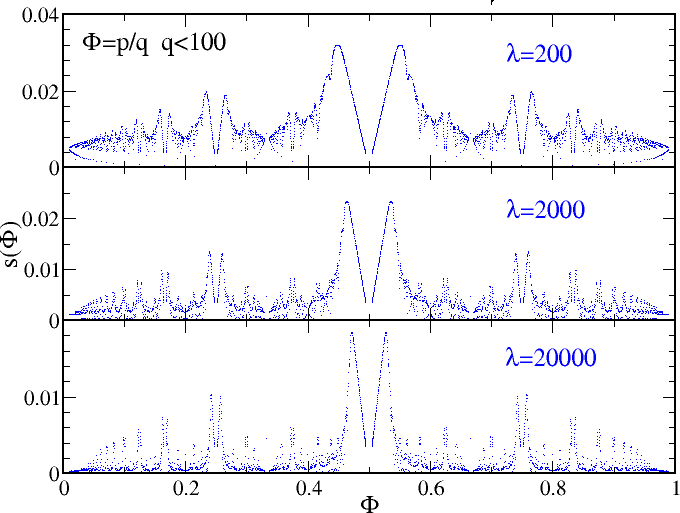}
\caption{The entropy $s(\lambda,\Phi)$ per unit cell as a function of
the magnetic flux $\Phi$ for $\lambda\gg1$. The figure contains all
fluxes $\Phi=p/q$ with $p$,$q$ coprime and $q<100$.}
\label{fig6}
\end{figure}
\begin{figure}
 \includegraphics[width=\columnwidth]{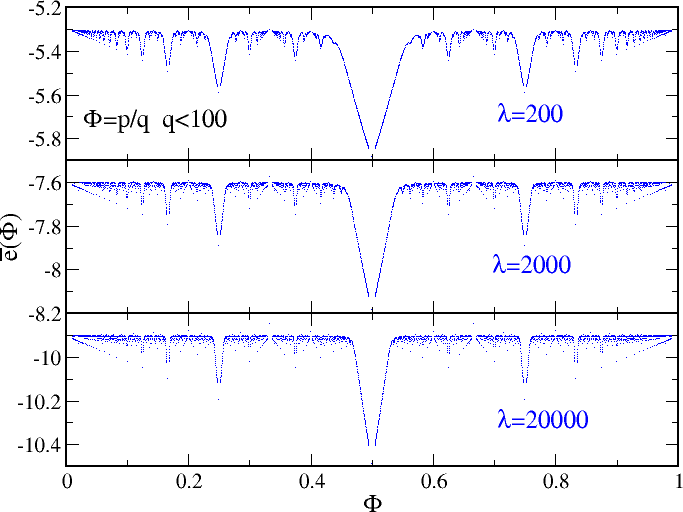}
\caption{The energy $\bar e(\lambda,\Phi)$ per unit cell defined by
Eq.~(\ref{e}) as a function of
the magnetic flux $\Phi$ for $\lambda\gg1$.}
\label{fig7}
\end{figure}

\section{Hofstadter Bilayers: Entanglement Thermodynamics}
\label{thermo}

In the spirit of Ref.~\cite{Schliemann11} we now investigate the
thermodynamics based on the entanglement Hamiltonian (\ref{entham})
and the pertaining reduced density operator $\rho_{\rm red}$. 
From Eq.~(\ref{ent2}) one finds
\begin{equation}
\xi_m(\vec k)=\lambda\tilde\varepsilon_m(\vec k)
-\frac{1}{24}\left(\lambda\tilde\varepsilon_m(\vec k)\right)^3
+{\cal O}\left(\lambda^5\right)
\end{equation}
implying (cf. Eq.~(\ref{HA})))
\begin{equation}
{\cal H}_{\rm ent}\approx\lambda{\cal H}_A/t_A
\label{entmono}
\end{equation}
for small coupling parameter $\lambda$. This suggests to interpret $\lambda$
as an inverse ``entanglement temperature'' 
and approximate the reduced density matrix as $\rho_{\rm red}\approx\rho_0$ with
\begin{equation}
\rho_0=\frac{1}{Z_0}\exp(-\lambda{\cal H}_A/t_A) 
\label{rho0}
\end{equation}
with $Z_0={\rm tr}(\exp(-\lambda{\cal H}_A/t_A))$ and 
${\cal H}_A/t_A$ describing energy of the subsystem.
The above
findings are completely analogous to the ones in Ref.~\cite{Schliemann11}
on quantum Hall bilayers where the above approximate relation
(\ref{entmono}), valid for strongly coupled layers, was used
to numerically 
analyze a quantum phase transition in the total double layer system.  
However, having a closed analytical result for the entanglement
levels at hand, we shall follow a different route here and employ
the full expressions (\ref{ent2}),(\ref{entham}) 
(not necessarily approximated by their first order in $\lambda$)
to evaluate thermodynamics in a more systematic manner.
\begin{figure}
 \includegraphics[width=\columnwidth]{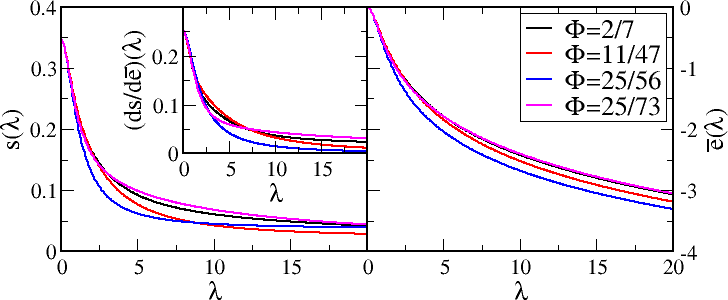}
\caption{Entropy $s(\lambda,\Phi)$ (left panel)
and energy $\bar e(\lambda,\Phi)$ (right panel) per unit
cell as a function of $\lambda$ for various fluxes $\Phi$.
The inset shows the derivative $\partial s/ \partial\bar e$. All quantities
decrease monotonously with increasing $\lambda$ and do not show
any apparent qualitative difference for different $\Phi$.}
\label{fig8}
\end{figure}
As we shall explore in more detail below, the coupling parameter $\lambda$
can be used as a phenomenological temperature scale, but it is not
identical to the inverse thermodynamic temperature given by the derivative
of the entropy with respect to the (appropriately defined) inner energy.

A very simple issue is the average particle number
$N=\langle\hat N\rangle$ with $\hat N=\sum_{m\vec k}a^+_{\vec k m}a_{\vec k m}$ and
$\langle\cdot\rangle={\rm tr}(\rho_{\rm red}\cdot)$. Here one always
has, independently of coupling parameter and magnetic flux,
$N(\lambda.\Phi)=N_xN_y/2$ which is clear from the fact that the
total bilayer system is half-filled and unbiased with respect to its subsystems.
Formally this result can be established, in the
thermodynamic limit $N_x,N_y\gg 1$, as follows,
\begin{eqnarray}
N(\lambda,\Phi) & = & N_xN_y
\frac{a^2}{(2\pi)^2}\sum_{m=0}^{q-1}\int_0^{2\pi/(qa)}dk_x\int_0^{2\pi/a}dk_y
\frac{1}{1+e^{\xi_m(\vec k)}}\nonumber\\
& = & N_xN_y
\frac{a^2}{(2\pi)^2}\sum_{m=0}^{q-1}\int_0^{2\pi/(qa)}dk_x\int_0^{2\pi/a}dk_y
\left(\frac{1}{2}+\frac{\lambda\tilde\varepsilon_m(\vec k)/4}
{\sqrt{1+(\lambda\tilde\varepsilon_m(\vec k)/2)^2}}\right)\nonumber\\
& = & N_xN_y/2\,,
\end{eqnarray}
where we have used Eqs.~(\ref{ent2a}) and (\ref{rel1}). 
This constancy of the averaged particle number can of course
also be seen as a consequence of the fact that a reduced density matrix
of the form (\ref{rhored}) can be viewed as a grand-canonical
statistical operator with constant chemical potential $\mu=0$, which lies
exactly in the middle of the symmetric entanglement spectrum.

Let us next turn to the other extensive quantities entropy and energy 
which we define, again in the thermodynamic limit, by
\begin{eqnarray}
S & = & \langle-\ln\rho_{\rm red}\rangle=N_xN_ys(\lambda,\Phi)\,,\\
\bar E & = & \langle H_{\rm ent}\rangle=N_xN_y\bar e(\lambda,\Phi)
\end{eqnarray}
with
\begin{eqnarray}
s(\lambda,\Phi) & = &
\frac{a^2}{(2\pi)^2}\sum_{m=0}^{q-1}\int_0^{2\pi/(qa)}dk_x\int_0^{2\pi/a}dk_y
\frac{\ln\left(1+e^{\xi_m(\vec k)}\right)}{1+e^{\xi_m(\vec k)}}\,,
\label{s}\\
\bar e(\lambda,\Phi) & = &
\frac{a^2}{(2\pi)^2}\sum_{m=0}^{q-1}\int_0^{2\pi/(qa)}dk_x\int_0^{2\pi/a}dk_y
\frac{\xi_m(\vec k)}{1+e^{\xi_m(\vec k)}}\,.
\label{e}\\
\end{eqnarray}
The bar at the above quantity $\bar E$ is meant to indicate that the
definition of energy will be subject to some refinement further below.
Moreover, as a consequence of Eq.~(\ref{rel3}), this quantity is non-positive,
\begin{equation}
\bar e(\lambda,\Phi)\leq0
\label{signe}
\end{equation}
for all $\lambda$ and $\Phi$.
Besides, since the entropy $S$ is obviously proportional
to the area of the system coinciding with the boundary to the
other subsystem, the so-called area law is fulfilled \cite{Eisert10}.

For small $\lambda\ll 1$ one obtains the expansions
\begin{eqnarray}
s(\lambda,\Phi) & = & \frac{1}{2}\ln 2-\frac{\lambda^2}{4}
+\lambda^4\frac{5}{96}\left(7+2\cos(2\pi\Phi)\right)
+{\cal O}\left(\lambda^6\right)
\,,\label{exps}\\
\bar e(\lambda,\Phi) & = &
-\lambda^2+\frac{\lambda^4}{6}\left(7+2\cos(2\pi\Phi)\right)
+{\cal O}\left(\lambda^6\right)\,,\label{expe}
\end{eqnarray}
where we have used the fact that 
\begin{eqnarray}
\omega_n(\Phi) & := & 
\frac{a^2}{(2\pi)^2}\sum_{m=0}^{q-1}\int_0^{2\pi/(qa)}dk_x\int_0^{2\pi/a}dk_y
\left(\tilde\varepsilon(\vec k)\right)^n\\
& = & 
\frac{a^2}{(2\pi)^2}\int_0^{2\pi/(qa)}dk_x\int_0^{2\pi/a}dk_y
{\rm tr}\left[\left(h(\vec k)\right)^n\right]
\end{eqnarray}
vanishes for odd $n$, and for low even values one has
$\omega_0(\Phi)\equiv1$, $\omega_2(\Phi)\equiv4$, and
\begin{equation}
\omega_4(\Phi)=4\left(7+2\cos(2\pi\Phi)\right),
\end{equation} 
as it is readily derived from the explicit form (\ref{hammat2}) of the
matrix $h(\vec k)$. We note that the zeroth-order value 
$s(0,\Phi)=(\ln 2)/2$ of the entropy per unit cell is due to
the fact that, at strong coupling, each particle ends up in either layer
with equal probability, and the total bilayer system is half-filled.

In Fig.~\ref{fig5} we have plotted the entropy $s(\lambda,\Phi)$ 
per unit cell as a function of 
the magnetic flux $\Phi$ for various coupling strength starting with
$\lambda$ being of order unity, where $s(\lambda,\Phi)$ is a smooth
function of $\Phi$. At values of  
$\lambda\approx10$ and larger, however, $s(\lambda,\Phi)$ becomes more
irregular and shows typical fractal features such as self-similarity.
As seen in Fig.~\ref{fig6}, this property of $s(\lambda,\Phi)$ becomes
more pronounced with increasing $\lambda$. A similar behavior
is found for $\bar e(\lambda,\Phi)$, as illustrated in Fig.~\ref{fig7}.

Fig.~\ref{fig8} shows the entropy $s(\lambda,\Phi)$ and the energy
$\bar e(\lambda,\Phi)$ as a function of $\lambda$ for a few representative
values of $\Phi$. As seen, both quantities decrease monotonically with
$\lambda$, and in this sense $\lambda$ qualifies as a phenomenological
(inverse) temperature scale. However, if $\lambda$ were the true inverse
thermodynamic temperature, 
standard thermodynamics would require it to equal the derivative
\begin{equation}
\frac{\partial s}{\partial\bar e}
=\frac{\partial s(\lambda,\Phi)}{\partial\lambda}
\left(\frac{\partial\bar e(\lambda,\Phi)}{\partial\lambda}\right)^{-1}\,,
\label{deriv1}
\end{equation}
where the derivatives with respect to $\lambda$ on the 
r.h.s can be expressed as
\begin{eqnarray}
\frac{\partial s(\lambda,\Phi)}{\partial\lambda} & = &
\frac{a^2}{(2\pi)^2}\sum_{m=0}^{q-1}\int_0^{2\pi/(qa)}dk_x\int_0^{2\pi/a}dk_y
\nonumber\\
 & & \cdot\frac{\left(-\ln\left(1+e^{\xi_m(\vec k)}\right)+1\right)
\tilde\varepsilon(\vec k)}
{\left(1+e^{\xi_m(\vec k)}\right)\left(1+e^{-\xi_m(\vec k)}\right)
\sqrt{1+(\lambda\tilde\varepsilon(\vec k)/2)^2}}\,,
\label{deriv2}\\
\frac{\partial\bar e(\lambda,\Phi)}{\partial\lambda} & = &
\frac{a^2}{(2\pi)^2}\sum_{m=0}^{q-1}\int_0^{2\pi/(qa)}dk_x\int_0^{2\pi/a}dk_y
\nonumber\\
 & & \cdot\frac{\left(-\xi_m(\vec k)+\left(1+e^{\xi_m(\vec k)}\right)\right)
\tilde\varepsilon(\vec k)}
{\left(1+e^{\xi_m(\vec k)}\right)\left(1+e^{-\xi_m(\vec k)}\right)
\sqrt{1+(\lambda\tilde\varepsilon(\vec k)/2)^2}}\,.
\label{deriv3}
\end{eqnarray} 
In particular, for small $\lambda\ll1$ one finds from 
Eqs.~(\ref{exps}) and (\ref{expe})
\begin{equation}
\frac{\partial s}{\partial\bar e}=\frac{1}{4}+{\cal O}\left(\lambda^2\right)\,,
\label{deriv4}
\end{equation}
independently of the magnetic flux $\Phi$. As seen in the inset of
Fig.~\ref{fig8}, and also very explicitly by Eq.~(\ref{deriv4}),
the derivative (\ref{deriv1}) is certainly not equal to $\lambda$.

The reason for this behavior is that the reduced density matrix
formulated as 
$\rho_{\rm red}=\exp(-{\cal H}_{\rm ent})/Z$ does not match a canonical
equilibrium state characterized by an inverse temperature $\beta$.
Let us therefore rewrite the entanglement Hamiltonian according to
\begin{equation}
{\cal H}_{\rm ent}(\lambda,\Phi)
=\beta(\lambda,\Phi){\cal H}_{\rm can}(\lambda,\Phi)\,,
\end{equation} 
where the inverse thermodynamic temperature $\beta(\lambda,\Phi)$
is determined as a function of $\lambda$ as follows: Defining the
thermodynamic inner energy 
\begin{equation}
E(\lambda,\Phi)=\langle{\cal H}_{\rm can}(\lambda,\Phi)\rangle
\end{equation}
and the free energy
\begin{eqnarray}
F(\lambda,\Phi) & = & E(\lambda,\Phi)-S(\lambda,\Phi)/ \beta(\lambda,\Phi)\\
 & = & -\ln(Z(\lambda,\Phi))/ \beta(\lambda,\Phi)
\end{eqnarray}
one easily verifies
\begin{equation}
\beta\frac{\partial\bar F}{\partial\beta}=\bar E
\label{Fbar}
\end{equation}
with $\bar F=\beta F=\bar E-S$, where we have used the stipulated relation
\begin{equation}
\frac{\partial S}{\partial E}=\beta\,.
\label{dS/dE}
\end{equation}
The above equations (\ref{Fbar}) and (\ref{dS/dE})
are indeed equivalent, and from Eq.~(\ref{Fbar}) it follows
\begin{equation}
\frac{\partial\ln\beta}{\partial\lambda}
=\frac{1}{\bar E}\frac{\partial\bar F}{\partial\lambda}
=\frac{1}{\bar e}\frac{\partial(\bar e-s)}{\partial\lambda}\,,
\label{betalambda1}
\end{equation}
which is the sought equation determining $\beta(\lambda)$.
For small $\lambda$ the expansions (\ref{exps}) and  (\ref{expe})
lead to
\begin{equation}
\frac{\partial\ln\beta}{\partial\lambda}
=\frac{3}{2\lambda}
+\frac{35}{12}\lambda(7+2\cos(2\pi\Phi))+{\cal O}\left(\lambda^2\right)
\label{betalambda2}
\end{equation}
such that
\begin{equation}
\ln\beta(\lambda,\Phi)
=\ln k+\frac{3}{2}\ln\lambda
+\frac{35}{24}\lambda^2(7+2\cos(2\pi\Phi))+{\cal O}\left(\lambda^3\right)\,.
\label{betalambda3}
\end{equation}
Here the integration constant $\ln k$, $k>0$, reflects the freedom of
choosing a unit to measure $\beta$, i.e. $k$ plays the same role as 
Boltzmann\rq s constant in standard thermodynamics. The entropy $s$ and 
energy $e=\bar e/ \beta$ per unit cell read as a function of small $\beta$
\begin{eqnarray}
s(\beta,\Phi) & = & \frac{1}{2}\ln 2-\frac{1}{4}\left(\beta/k\right)^{4/3}
+{\cal O}\left(\beta^3\right)\,,\\
e(\beta,\Phi) & = & -\frac{1}{k}\left(\beta/k\right)^{1/3}
+{\cal O}\left(\beta^2\right)\,.
\end{eqnarray}

Remarkably, the inverse thermodynamic temperature 
\begin{equation}
\beta(\lambda)=k\lambda^{3/2}+{\cal O}\left(\lambda^{7/2}\right)
\label{betalambda4}
\end{equation}
scales for $\lambda\ll 1$ as $\lambda^{3/2}$ and is {\em not} linear
in this parameter, differently what would follow from the
{\em ansatz} $\rho_{\rm red}=\rho_0$, cf. Eq.~(\ref{rho0}). The technical reason
for this observation is that $\rho_0$ is, while being of canonical form,
{\em not} the linear approximation to $\rho_{\rm red}$ but contains also
arbitrary high powers in $\lambda$. 

At large $\lambda$, i.e. weak coupling between the layers, the entropy $s$ 
characterizing their mutual entanglement will 
(along with its derivative) eventually vanish.
Thus, for $\lambda\gg1$ Eq.~(\ref{betalambda1}) can be simplified as
\begin{equation}
\frac{\partial\ln\beta}{\partial\lambda}
\approx\frac{\partial\ln|\bar e|}{\partial\lambda}
\label{betalambda5}
\end{equation} 
leading to
\begin{equation}
\beta(\lambda,\Phi)\approx k|\bar e(\lambda,\Phi)|=-k\bar e(\lambda,\Phi)
\label{betalambda6}
\end{equation} 
where we have used Eq.~(\ref{signe}) and have adjusted the
constant $k$ consistent with Eq.~(\ref{betalambda3}). In particular,
the above relation implies
\begin{equation}
e(\lambda,\Phi)\approx-\frac{1}{k}
\label{betalambda7}
\end{equation}
for large $\lambda$. We note that the sign on the r.h.s. of
Eqs.~(\ref{betalambda7}) depends on the
inequality (\ref{signe}) and would be different for positive 
$\bar e(\lambda,\Phi)$ at $\lambda\gg1$.
\begin{figure}
 \includegraphics[width=\columnwidth]{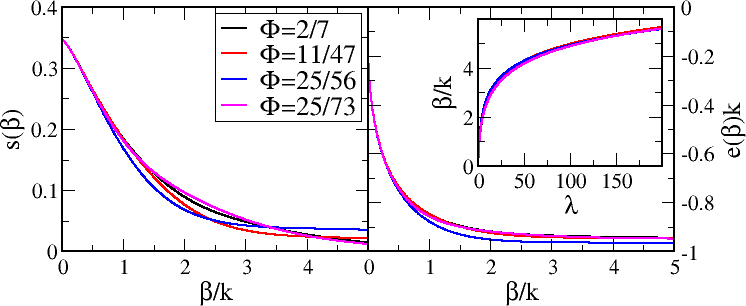}
\caption{Entropy $s$ (left panel)
and energy $e$ (right panel) per unit
cell as a function of the inverse thermodynamic temperature
$\beta$ for the same flux values $\Phi$ as in Fig.~\ref{fig8}.
The inset shows $\beta$ as a function of $\lambda$. This data was obtained
via a numerical integration of Eq.~(\ref{betalambda1}) starting from 
expression (\ref{betalambda3}) with an appropriately small 
but finite initial argument $0<\lambda\ll1$.}
\label{fig9}
\end{figure}

\begin{figure}
\includegraphics[width=\columnwidth]{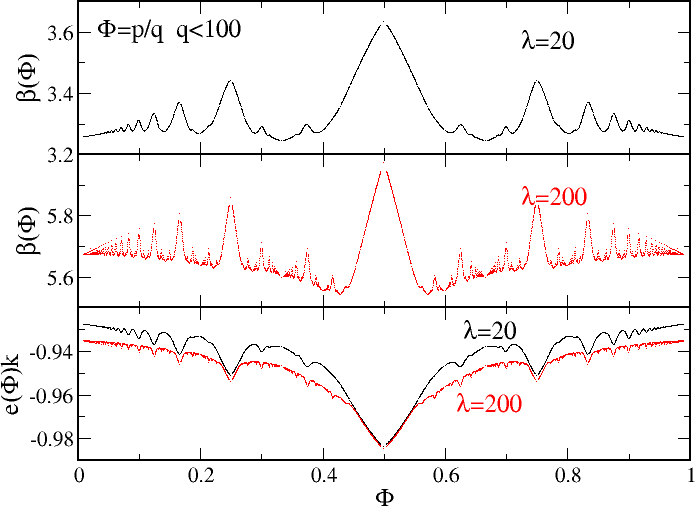}
\caption{The inverse thermodynamic temperature $\beta(\lambda,\Phi)$
(upper two panels) and the inner energy  $\e(\lambda,\Phi)$ per unit cell
(bottom panel)
as a function of the magnetic flux $\Phi$ for large $\lambda\gg1$.
The figure contains all
fluxes $\Phi=p/q$ with $p$,$q$ coprime and $q<100$.}
\label{fig10}
\end{figure}

The inset of Fig.~\ref{fig9} shows $\beta$ as a function of $\lambda$
obtained via numerical integration of Eq.~(\ref{betalambda1}) 
for the same flux values $\Phi$ as in Fig.~\ref{fig8}.
The data depends only weakly on $\Phi$ and 
follows the power law (\ref{betalambda4}) at small $\lambda$ 
while the behavior at large $\lambda$ is well described by a logarithmic
dependence. In the main panels of Fig.~\ref{fig9} we have plotted
the entropy $s(\beta,\Phi)$ and the energy $e(\beta,\Phi)$ (as opposed
to $\bar e$) as a function of $\beta$. for large $\beta$, $e(\beta,\Phi)$
levels off and converges (slowly) to  $e=-1/k$,
consistent with Eq.~(\ref{betalambda7}).
Such a saturation of the thermodynamic energy should
be expected since both layers get more and more decoupled in this limit.
The originally defined energy $\bar e=\beta e$ does, in accordance
with Eq.~(\ref{betalambda6}), not show
such a behavior but decreases unboundedly with increasing $\lambda$,
as shown already in Figs.~\ref{fig7} and \ref{fig8}.
While approaching their ground state values in the limit
$\lambda\to\infty$ the inverse thermodynamic temperature $\beta(\lambda,\Phi)$
as well as the inner energy $\e(\lambda,\Phi)$ per unit cell
display fractal features,
as illustrated in Fig.~\ref{fig10}. An analogous behavior was found for
the entropy in Fig.~\ref{fig6}.

In summary, the parameter $\lambda$ can be viewed as a phenomenological
inverse temperature scale having intuitive properties like the
decrease of energy and entropy with increasing $\lambda$. The thermodynamic
inverse temperature fulfilling standard thermodynamic relations, 
however, is given by $\beta(\lambda)$. Eq.~(\ref{betalambda1}) establishes
a $1$-to-$1$ mapping between both quantities and is completely
universal in the sense that its derivation does not depend on  any
detail of the underlying system or its entanglement Hamiltonian.
Indeed, a relation very similar
to Eq.~(\ref{betalambda1}) (with in fact an identical l.h.s.) is obtained
in standard thermodynamics when connecting the thermodynamic temperature
to a phenomenological temperature scale obtained from, say, the
change in volume of a given body upon changing its internal energy
\cite{Landau80,Schwabl06}. Also the results (\ref{betalambda5}) to 
(\ref{betalambda7}) are (up to the sign (\ref{signe}) of $\bar e$)
very general since they only rely on the vanishing
of the derivative of the entanglement entropy 
$s(\lambda)$ in the limit of weak coupling
$\lambda\gg1$ ($\Leftrightarrow\beta\gg1$). This statement is of course
just the analog of the third law of thermodynamics which requires the
entropy to approach a constant
in the limit of zero temperature. In the same limit, the thermodynamic energy
$e$ per unit cell reaches according to  Eq.~(\ref{betalambda7})
a saturation corresponding to the ground state energy in classic
thermodynamics. Remarkably, within the formalism of entanglement
thermodynamics outlined here, this limit is universal and depends
only on the constant $k$, i.e. the unit chosen to measure temperature. 

\section{Conclusions and Outlook}
\label{concl}

We have derived an explicit expression for the entanglement levels
of Hofstadter bilayers in terms of the energy eigenvalues of the underlying 
monolayer system. 
For strongly coupled layers we find the (expected) proportionality between
the entanglement Hamiltonian and the energetic Hamiltonian of the
monolayer system with the proportionality factor given by an effective
coupling parameter. This parameter, however, is not identical to
the inverse thermodynamic temperature, but represents a
phenomenological temperature scale. We have devised an explicit relation between
both temperature scales which is in close analogy to a standard result of
classic thermodynamics. The introduction of the thermodynamic
temperature also implies a redefinition of the inner energy.
In the limit of vanishing temperature, thermodynamic
quantities such as entropy and inner energy approach their ground-state
values, but show a fractal structure as a function of magnetic flux.

The relation between the phenomenological temperature scale
(given by an appropriate coupling parameter of the total system)
and the thermodynamic entanglement temperature applies certainly also
to other and more general situations. For instance, in Ref.~\cite{Schliemann11}
an entanglement temperature was obtained for bilayer quantum Hall systems
by a numerical fit to exact-diagonalization data as a function of layer
separation. In the light of the present work, this temperature scale should
be seen as a phenomenological one. The analysis of the quantum phase transition
performed there, however, should qualitatively not be affected by this
issue, since the relation between a phenomenological and the thermodynamic
temperature is expected to be a smooth function. On the other and, very explicit
investigations as done here for non-interacting particles are
obviously more difficult for an interacting system since clearly less
analytical tools are available \cite{Schliemann11}.

Similar comments apply to the situation of spin ladders in
the limit of strong rung coupling \cite{Peschel11,Lauchli11,Schliemann12}:
The proportionality factor found there between entanglement spectrum and
energy spectrum should also be viewed as a phenomenological temperature
scale, but not the thermodynamic one. Moreover, Ref.~\cite{Eisler06} also
introduces, studying block entanglement in a spin-$1/2$ $XX$-chain, an
effective inverse temperature which grows monotonically with increasing block 
size. Although the situation there and the present one show obvious differences
their possible interrelations are of interest.

Finally, another possible extension of the present study is to consider
other types of lattices. Given the large deal of interest devoted presently
to graphene, the hexagonal geometry \cite{Haegawa06}
and its bilayer versions \cite{Nemec07} are obvious candidates . Indeed, it is
an interesting speculation whether, for example, the exponent $3/2$ 
occuring in Eq.~(\ref{betalambda4}) depends on the lattice geometry. 

\section*{Acknowledgements}

I thank J. Carlos Egues and Esmerindo S. Bernardes for earlier joint
work on Hofstadter systems.
This work was supported by DFG via SFB 631.

{}

\end{document}